\documentclass[10pt, conference]{IEEEtran}
\IEEEoverridecommandlockouts
\usepackage{cite}
\usepackage{amsmath, amssymb, amsfonts}
\usepackage{algorithmic}
\usepackage{graphicx}
\usepackage{textcomp}
\usepackage{xcolor}
\usepackage{booktabs}
\usepackage{multirow}
\usepackage{array}
\usepackage{makecell}
\usepackage{longtable}
\usepackage{subcaption}
\usepackage{tikz}
\usetikzlibrary{positioning, arrows.meta}
\usepackage{pgfplots}
\pgfplotsset{compat=1.18}
\usepackage{placeins}
\usepackage{dblfloatfix}
\usepackage{url}
\usepackage{hyperref}
\usepackage{tabularx}
\usepackage{algorithm}

\def\BibTeX{{\rm B\kern-.05em{\sc i\kern-.025em b}\kern-.08em T\kern-.1667em\lower.7ex\hbox{E}\kern-.125emX}}

\begin{document}
    \title{Progressive Per-Branch Depth Optimization for DEFOM-Stereo and SAM3
    Joint Analysis in UAV Forestry Applications}
    \author{\IEEEauthorblockN{Yida Lin, Bing Xue, Mengjie Zhang} \IEEEauthorblockA{\small \textit{Centre for Data Science and Artificial Intelligence} \\ \textit{Victoria University of Wellington, Wellington, New Zealand}\\ linyida\texttt{@}myvuw.ac.nz, bing.xue\texttt{@}vuw.ac.nz, mengjie.zhang\texttt{@}vuw.ac.nz}
    \and \IEEEauthorblockN{Sam Schofield, Richard Green} \IEEEauthorblockA{\small \textit{Department of Computer Science and Software Engineering} \\ \textit{University of Canterbury, Canterbury, New Zealand}\\ sam.schofield\texttt{@}canterbury.ac.nz, richard.green\texttt{@}canterbury.ac.nz}
    }
    \maketitle
    \vspace{-1.5em}

    \begin{abstract}
        Accurate per-branch 3D reconstruction is a prerequisite for autonomous UAV-based
        tree pruning, yet dense disparity maps from modern stereo matchers
        remain too noisy for individual branch analysis in complex forest
        canopies. This paper introduces a progressive pipeline that chains DEFOM-Stereo
        foundation-model disparity estimation, SAM3 instance segmentation, and multi-stage
        depth optimization to deliver robust per-branch point clouds. Beginning
        with a na\"ive baseline (Version~1), we identify and resolve three error
        families across six successive refinements. Mask boundary contamination
        is first tackled through morphological erosion (Version~2) and
        subsequently through a skeleton-preserving variant that safeguards thin-branch
        topology (Version~3). Segmentation inaccuracy is then addressed with LAB-space
        Mahalanobis color validation coupled with cross-branch overlap arbitration
        (Version~4). Depth noise---the final and most persistent error source---is
        initially mitigated by IQR/Z-score outlier removal and median filtering
        (Version~5), before being superseded by a five-stage robust scheme: MAD
        global detection, spatial density consensus, local MAD filtering, RGB-guided
        filtering, and adaptive bilateral filtering (Version~6). Evaluated on $19
        20{\times}1080$ stereo imagery of Radiata pine acquired with a ZED~Mini camera
        (63~mm baseline) from a UAV in Canterbury, New Zealand, the final pipeline
        reduces average per-branch depth standard deviation by 82\% while
        retaining edge fidelity---yielding geometrically coherent 3D point
        clouds ready for autonomous pruning tool positioning. All code and processed
        outputs are publicly released to facilitate further UAV forestry research.
    \end{abstract}

    \vspace{-1em}

    \section{Introduction}

    Radiata pine (\textit{Pinus radiata}) dominates New Zealand's plantation
    landscape, underpinning a forestry sector worth NZ\$3.6~billion (1.3\% of GDP).
    Producing high-value clear wood demands regular pruning, yet the manual
    process exposes workers to serious hazards---falls and chainsaw injuries
    chief among them. Autonomous UAV-based pruning promises a safer pathway~\cite{lin2024branch,steininger2025timbervision},
    but its realization hinges on centimeter-level depth knowledge \emph{per
    individual branch} at 1--2~m operating distances.

    The convergence of two recent breakthroughs now makes per-branch 3D
    reconstruction from drone stereo imagery a practical prospect. On the depth side,
    DEFOM-Stereo~\cite{jiang2025defom}---a foundation model coupling a DINOv2~\cite{oquab2024dinov2}
    ViT-L encoder with iterative stereo decoding---delivers state-of-the-art
    cross-domain generalization, producing dense disparity maps on unseen vegetation
    scenes without fine-tuning~\cite{lin2024benchmark}. On the segmentation side,
    the Segment Anything Model 3 (SAM3)~\cite{ravi2024sam2} provides class-agnostic
    instance masks that reliably delineate individual branches despite heavy occlusion.
    Chaining these two components opens the door to extracting per-branch depth
    maps and reconstructing 3D point clouds---a capability that neither stereo-only
    nor segmentation-only pipelines can offer alone.

    In practice, however, a straightforward fusion of DEFOM and SAM3 exposes
    three interacting error sources that severely compromise per-branch 3D
    fidelity. First, SAM3 masks tend to extend slightly past true branch
    contours, admitting background (sky) pixels with drastically different depth
    values---what we refer to as \textbf{mask boundary contamination}. Second,
    confidence-based filtering is insufficient to prevent color-inconsistent
    pixels from infiltrating branch masks or to disentangle overlapping masks
    between neighboring branches (\textbf{segmentation inaccuracy}). Third, DEFOM
    disparity maps, although globally accurate, harbor per-pixel noise,
    spatially isolated outliers, and boundary artifacts that corrupt per-branch depth
    histograms and 3D reconstructions (\textbf{depth noise}).

    Rather than attempting a monolithic solution, we adopt a \emph{progressive}
    strategy: six successive pipeline versions, each diagnosing and correcting a
    specific failure mode surfaced by its predecessor.

    \begin{itemize}
        \item \textbf{Version~1 (Baseline)} directly chains DEFOM disparity
            estimation, depth conversion, SAM3 segmentation, and per-branch point
            cloud generation; no refinement is applied.

        \item \textbf{Versions~2--3} attack mask boundary contamination---first
            with standard morphological erosion (V2), then with a skeleton-preserving
            variant that maintains thin-branch connectivity via distance transforms
            and topological skeletonization (V3).

        \item \textbf{Version~4} targets segmentation inaccuracy through a four-stage
            refinement sequence: boundary erosion with skeleton connectivity,
            LAB Mahalanobis color validation, connected-component cleaning, and
            cross-branch overlap arbitration.

        \item \textbf{Version~5} addresses depth noise via multi-round IQR
            outlier removal, Z-score filtering, local spatial detection, and median
            smoothing.

        \item \textbf{Version~6 (Final)} supersedes V5 with a more robust five-stage
            depth pipeline: MAD global outlier detection, spatial density
            consensus, local MAD filtering, RGB-guided filtering~\cite{he2013guided},
            and adaptive bilateral filtering.
    \end{itemize}

    The principal contributions of this work are as follows:

    \begin{itemize}
        \item The first end-to-end pipeline uniting a stereo foundation model (DEFOM)
            with instance segmentation (SAM3) for per-branch 3D reconstruction in
            forestry.

        \item A skeleton-preserving mask erosion algorithm that protects thin-branch
            topology during boundary refinement.

        \item A LAB Mahalanobis color-validation scheme enabling pixel-level segmentation
            verification and cross-branch overlap resolution.

        \item A five-stage robust depth optimization pipeline that integrates MAD
            statistics, spatial consensus, guided filtering, and adaptive
            bilateral filtering for edge-preserving depth denoising.

        \item A systematic ablation study quantifying each refinement stage's contribution
            to final per-branch 3D quality.
    \end{itemize}

    \section{Related Work}

    \subsection{Deep Stereo Matching and Foundation Models}

    End-to-end trainable architectures have reshaped stereo matching over the
    past decade. Spatial pyramid pooling for multi-scale cost aggregation (PSMNet~\cite{chang2018psmnet}),
    group-wise correlation volumes (GwcNet~\cite{guo2019gwcnet}), and iterative refinement
    borrowed from optical flow (RAFT-Stereo~\cite{lipson2021raft}) represent successive
    leaps in disparity quality. A more recent paradigm shift towards \emph{foundation
    models} prioritizes cross-domain robustness over benchmark-specific accuracy.
    DEFOM-Stereo~\cite{jiang2025defom} exemplifies this trend: by pairing a DINOv2~\cite{oquab2024dinov2}
    ViT-L encoder with iterative stereo decoding, it achieves state-of-the-art
    zero-shot performance across ETH3D~\cite{schops2017eth3d}, KITTI~\cite{geiger2012kitti},
    and Middlebury~\cite{scharstein2014middlebury}---delivering consistent
    disparity maps with no catastrophic failures. Its suitability for vegetation
    scenes was confirmed in our earlier benchmark study~\cite{lin2024benchmark},
    which motivates its adoption as the depth front-end here.

    \subsection{Instance Segmentation for Vegetation}

    Promptable, class-agnostic segmentation became widely accessible through the
    Segment Anything Model (SAM)~\cite{kirillov2023sam}. Its successor, SAM~2~\cite{ravi2024sam2},
    introduced memory-augmented transformers for video-level reasoning, while
    SAM3 further sharpens mask boundaries through multi-scale feature fusion. Forestry
    applications have thus far concentrated on individual tree crown delineation~\cite{lin2025segmentation};
    \emph{per-branch} segmentation and the downstream depth extraction it enables
    remain largely uncharted territory.

    \subsection{Depth Map Refinement and Filtering}

    Stereo disparity maps are well known to suffer from noise, outliers, and
    edge bleeding. Bilateral filtering~\cite{tomasi1998bilateral}, guided
    filtering~\cite{he2013guided}, and weighted median filtering~\cite{ma2013constant}
    constitute the classical arsenal for alleviating these artifacts. On the
    statistical front, IQR- and MAD-based outlier detection~\cite{leys2013detecting}
    is standard practice in robust statistics, yet its spatially adaptive, per-instance
    application to stereo depth remains uncommon. Our pipeline traces an evolution
    from basic IQR clipping to a composite five-stage strategy that weaves together
    global MAD, local spatial consensus, guided filtering, and adaptive
    bilateral filtering.

    \subsection{Stereo Vision for UAV Forestry}

    Drone-mounted stereo cameras have been deployed for navigation~\cite{fraundorfer2012vision},
    3D reconstruction~\cite{nex2014uav}, and obstacle avoidance~\cite{barry2015pushbroom}.
    More closely aligned with our goals, recent efforts in branch detection~\cite{lin2024branch,lin2025yolosgbm}
    and segmentation~\cite{lin2025segmentation} illustrate how deep learning and
    stereo vision can jointly serve forestry needs. What remains missing is a
    pipeline that fuses foundation-model stereo with instance segmentation for
    per-branch 3D point cloud generation. The present work fills precisely this gap,
    offering a complete pathway from raw stereo pairs to optimized per-branch
    point clouds.

    \section{Methodology}

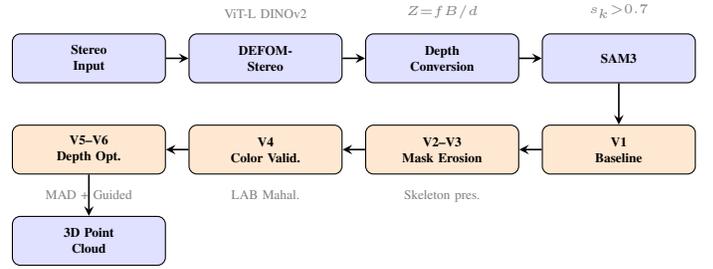
\begin{figure}[!t]
        \centering
        \begin{tikzpicture}[
            node distance=0.45cm and 0.3cm,
            box/.style={draw, rounded corners=3pt, minimum height=0.65cm, text width=1.8cm, align=center, font=\tiny\bfseries},
            stage/.style={box, fill=blue!12},
            ver/.style={box, fill=orange!18},
            arr/.style={->, >=stealth, semithick},
            lbl/.style={font=\tiny, text=gray},
        ]
            \node[stage] (stereo) {Stereo\\Input};
            \node[stage, right=of stereo] (defom) {DEFOM-\\Stereo};
            \node[stage, right=of defom] (depth) {Depth\\Conversion};
            \node[stage, right=of depth] (sam) {SAM3};
            \draw[arr] (stereo) -- (defom);
            \draw[arr] (defom) -- (depth);
            \draw[arr] (depth) -- (sam);
            \node[lbl, above=0.08cm of defom] {ViT-L DINOv2};
            \node[lbl, above=0.08cm of depth] {$Z{=}fB/d$};
            \node[lbl, above=0.08cm of sam] {$s_{k}{>}0.7$};
            \node[ver, below=0.55cm of sam] (v1) {V1\\Baseline};
            \draw[arr] (sam) -- (v1);
            \node[ver, left=of v1] (v23) {V2--V3\\Mask Erosion};
            \node[ver, left=of v23] (v4) {V4\\Color Valid.};
            \node[ver, left=of v4] (v56) {V5--V6\\Depth Opt.};
            \draw[arr] (v1) -- (v23);
            \draw[arr] (v23) -- (v4);
            \draw[arr] (v4) -- (v56);
            \node[lbl, below=0.08cm of v23] {Skeleton pres.};
            \node[lbl, below=0.08cm of v4] {LAB Mahal.};
            \node[lbl, below=0.08cm of v56] {MAD + Guided};
            \node[stage, below=0.55cm of v56] (pc) {3D Point\\Cloud};
            \draw[arr] (v56) -- (pc);
        \end{tikzpicture}
        \caption{Overview of the progressive pipeline. Foundation modules (top, blue)
        feed into SAM3; the arrow descends to Version~1, then proceeds right-to-left
        through six iterative refinements (orange) addressing mask contamination
        (V2--V3), segmentation accuracy (V4), and depth noise (V5--V6), producing
        per-branch 3D point clouds.}
        \label{fig:pipeline_overview}
    \end{figure}

    An overview of the complete pipeline appears in Fig.~\ref{fig:pipeline_overview}.
    The subsections below first lay out the shared foundation modules (\S\ref{sec:foundation})
    and then walk through each refinement stage in the order it was developed (\S\ref{sec:v1}--\S\ref{sec:v6}).

    \subsection{Shared Foundation}
    \label{sec:foundation}

    \subsubsection{Data Acquisition}
\begin{figure}[!t]
        \centering
        \begin{subfigure}
            [b]{0.48\columnwidth}
            \includegraphics[width=\textwidth]{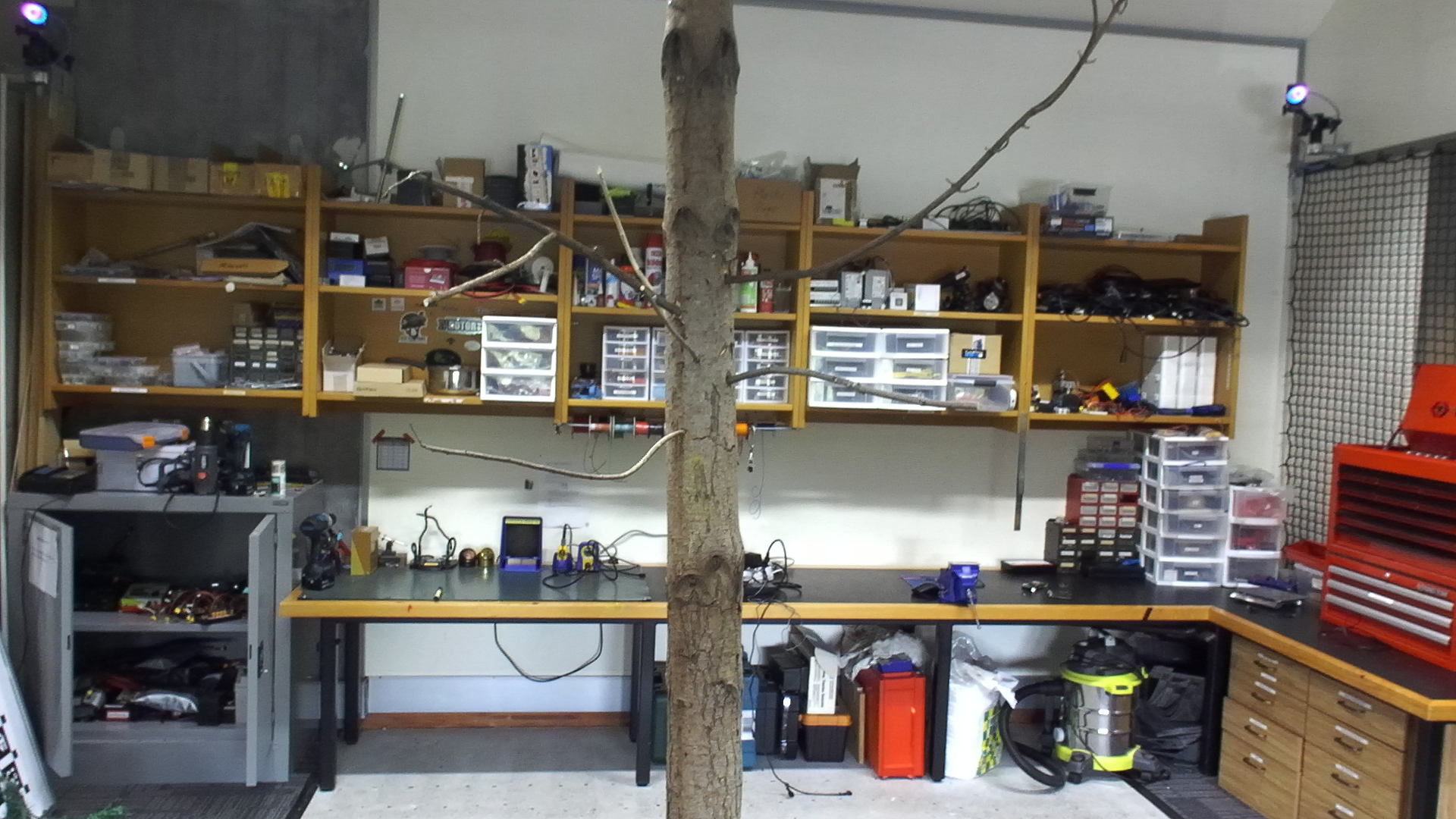}
            \caption{Left stereo image}
        \end{subfigure}
        \hfill
        \begin{subfigure}
            [b]{0.48\columnwidth}
            \includegraphics[width=\textwidth]{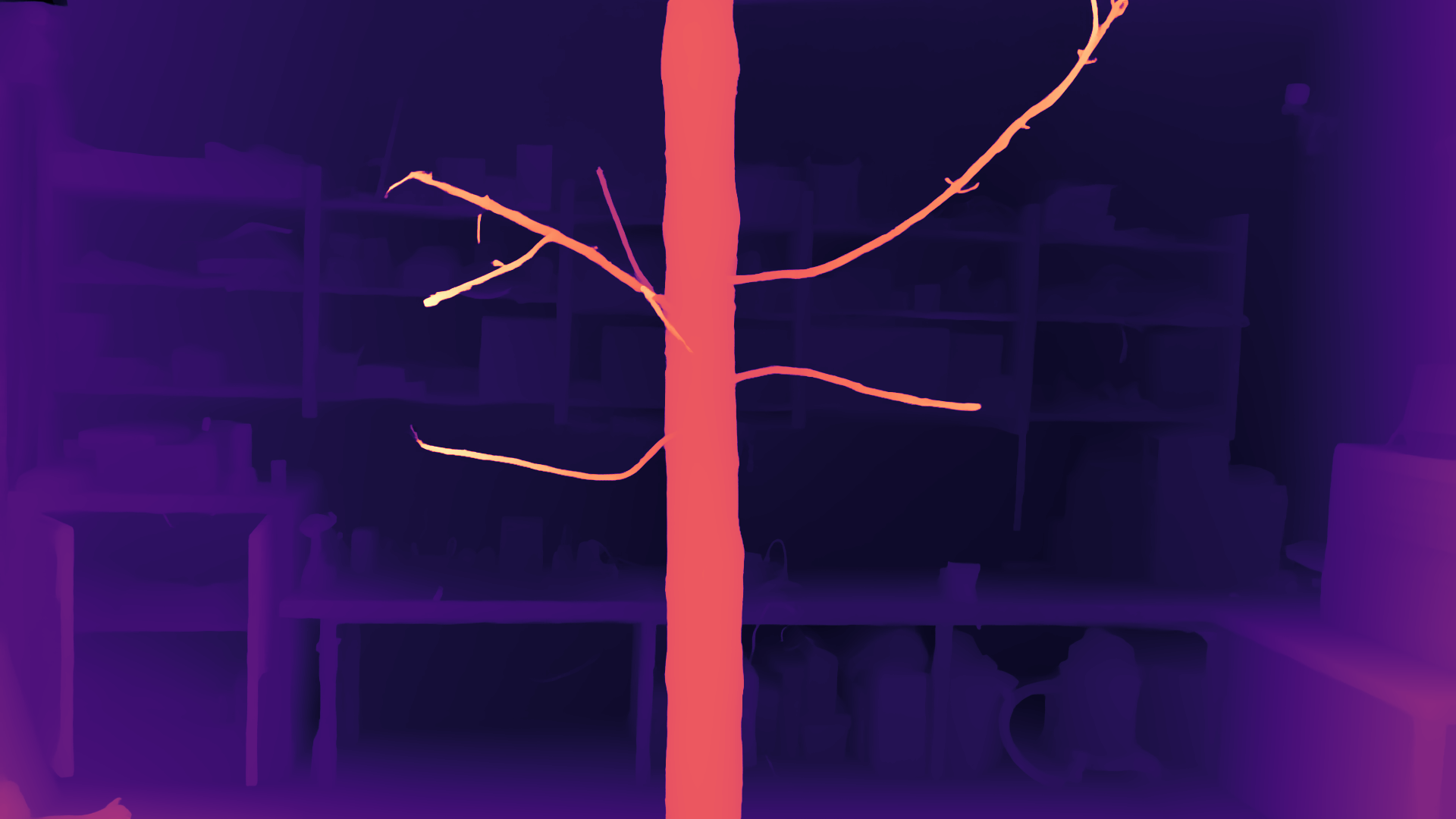}
            \caption{DEFOM disparity}
        \end{subfigure}
        \caption{Representative input data: (a)~left image of Radiata pine canopy
        captured by ZED~Mini at 1--2~m, and (b)~dense disparity map produced by DEFOM-Stereo
        (ViT-L DINOv2, 32 iterations). Warm colors indicate closer objects.}
        \label{fig:input_data}
    \end{figure}

    We capture stereo image pairs at $1920 \times 1080$ resolution using a ZED~Mini
    camera ($f_{x}\approx 1120$~px, baseline $B = 63$~mm) mounted on a UAV. Images
    are captured at 1--2~m distance from Radiata pine branches in Canterbury, New
    Zealand, under varying illumination conditions (March--October 2024). Fig.~\ref{fig:input_data}
    shows a representative left image and the corresponding DEFOM-Stereo disparity
    map.

    \subsubsection{DEFOM-Stereo Disparity Estimation}
    Given a rectified stereo pair
    $(I_{L}, I_{R}) \in \mathbb{R}^{H \times W \times 3}$, DEFOM-Stereo produces
    a dense disparity map $D \in \mathbb{R}^{H \times W}$. The model uses a
    DINOv2 ViT-L encoder with 32 valid iterations and 8 scale iterations. From disparity,
    depth is recovered by triangulation:
    \begin{equation}
        Z(x,y) = \frac{f_{x}\cdot B}{D(x,y)}\label{eq:depth}
    \end{equation}
    where $f_{x}$ is the horizontal focal length and $B$ is the stereo baseline.

    \subsubsection{3D Point Cloud Generation}
    Each pixel $(u, v)$ with valid depth $Z$ is back-projected to 3D using the standard
    pinhole model:
    \begin{equation}
        X = \frac{(u - c_{x}) \cdot Z}{f_{x}}, \quad Y = \frac{(v - c_{y}) \cdot
        Z}{f_{y}}\label{eq:backproject}
    \end{equation}
    where $(c_{x}, c_{y})$ are the principal point coordinates. The resulting point
    cloud $\mathcal{P}= \{(X_{i}, Y_{i}, Z_{i}, R_{i}, G_{i}, B_{i})\}$ preserves
    RGB color from the left image.

    \subsubsection{SAM3 Instance Segmentation}
    SAM3 is applied to the left image to produce a set of branch masks $\{M_{k}\}
    _{k=1}^{K}$ with confidence scores $\{s_{k}\}$. Only masks with $s_{k}> 0.7$
    are retained. For each accepted mask $M_{k}$, the per-branch depth map is
    $Z_{k}= Z \odot M_{k}$ (element-wise masking).

    \subsection{Version~1: Baseline Pipeline}
    \label{sec:v1}

    In its simplest form the pipeline is a straightforward cascade: DEFOM
    disparity $\rightarrow$ depth conversion $\rightarrow$ SAM3 segmentation
    $\rightarrow$ per-branch depth extraction and point cloud generation, with no
    post-processing of any kind.

    Although this baseline is functionally complete, it is afflicted by all
    three error families simultaneously: sky-depth outliers bleed into branch histograms
    through imprecise mask boundaries, color-inconsistent pixels pollute SAM3
    masks, and raw DEFOM noise scatters points across the 3D reconstructions.

    \subsection{Version~2: Morphological Mask Erosion}
    \label{sec:v2}

    The most immediate remedy for boundary contamination is to shrink each SAM3
    mask inward before extracting depth. Version~2 accomplishes this with morphological
    erosion using an elliptical structuring element of radius $r_{e}= 15$~px:
    \begin{equation}
        M_{k}^{\text{eroded}}= M_{k}\ominus \mathcal{E}_{r_e}\label{eq:erosion}
    \end{equation}
    where $\ominus$ denotes morphological erosion and $\mathcal{E}_{r_e}$ is the
    structuring element. A fallback mechanism progressively reduces $r_{e}$ to $\{
    r_{e}/2,\; r_{e}/4,\; 3,\; 1\}$ if the erosion destroys the mask entirely (i.e.,
    $|M_{k}^{\text{eroded}}| = 0$).

    \textbf{Limitation}: Standard morphological erosion works well for thick branches
    but disconnects or entirely removes thin branches whose diameter falls below
    $2r_{e}$---a shortcoming that Version~3 specifically targets.

    \subsection{Version~3: Skeleton-Preserving Erosion}
    \label{sec:v3}

    To retain thin-branch connectivity, Version~3 introduces a topology-aware erosion
    strategy that fuses distance-transform erosion with topological skeletonization:

    \textbf{Step~1: Distance transform erosion}. Compute the Euclidean distance transform
    $\mathcal{D}(x,y)$ of the binary mask $M_{k}$, yielding the distance from
    each foreground pixel to the nearest boundary. Retain only pixels sufficiently
    far from the boundary:
    \begin{equation}
        M_{k}^{\text{dist}}= \{(x,y) : \mathcal{D}(x,y) \geq r_{e}\} \label{eq:dist_erosion}
    \end{equation}

    \textbf{Step~2: Topological skeletonization}. Extract the morphological skeleton
    $\mathcal{S}(M_{k})$ of the original mask using the algorithm from~\cite{lee1994skeleton}.
    The skeleton is a one-pixel-wide connected centerline that preserves the
    topology of the mask.

    \textbf{Step~3: Skeleton dilation}. Dilate the skeleton with a small radius
    $r_{s}= \max(3, r_{e}/4)$ to give it pixel width:
    \begin{equation}
        \mathcal{S}^{\text{wide}}_{k}= \mathcal{S}(M_{k}) \oplus \mathcal{E}_{r_s}
        \label{eq:skel_dilate}
    \end{equation}

    \textbf{Step~4: Union}. The final refined mask is the union of the distance-eroded
    region and the widened skeleton, clipped to the original mask:
    \begin{equation}
        M_{k}^{\text{skel}}= (M_{k}^{\text{dist}}\cup \mathcal{S}^{\text{wide}}_{k}
        ) \cap M_{k}\label{eq:skel_union}
    \end{equation}

    Because the skeleton bridge persists even when $M_{k}^{\text{dist}}$ is empty
    for a narrow section, thin branches are guaranteed to remain topologically
    connected.

    \subsection{Version~4: Color-Validated Segmentation}
    \label{sec:v4}

    Skeleton-preserving erosion resolves boundary geometry, yet SAM3 masks may
    still contain color-inconsistent pixels---foliage fragments attached to a
    branch mask, for instance---and neighboring masks can overlap. Version~4
    tackles both issues through a four-stage segmentation refinement pipeline.

    \subsubsection{Stage~1: Boundary Erosion with Skeleton Preservation}
    Identical to Version~3 (Eq.~\ref{eq:dist_erosion}--\ref{eq:skel_union}).

    \subsubsection{Stage~2: LAB Mahalanobis Color Validation}

    For each branch mask $M_{k}$, we build a color model from the deeply-eroded
    \emph{core region} $C_{k}$ and reject pixels whose color deviates significantly.

    \textbf{Core region extraction}: Compute the distance transform of $M_{k}$ and
    retain only inner pixels with distance $\geq r_{c}= 25$~px:
    \begin{equation}
        C_{k}= \{(x,y) : \mathcal{D}_{M_k}(x,y) \geq r_{c}\} \label{eq:core}
    \end{equation}
    If $|C_{k}| < 100$, $r_{c}$ is progressively reduced until a sufficient core
    is obtained.

    \textbf{Gaussian color model in CIELAB space}: Convert the left image to CIELAB
    and compute the mean $\boldsymbol{\mu}_{k}$ and covariance $\boldsymbol{\Sigma}
    _{k}$ of core-region pixels:
    \begin{align}
        \boldsymbol{\mu}_{k}    & = \frac{1}{|C_{k}|}\sum_{(x,y) \in C_k}\mathbf{L}(x,y) \label{eq:color_mean}       \\
        \boldsymbol{\Sigma}_{k} & = \text{Cov}\bigl(\{\mathbf{L}(x,y)\}_{(x,y) \in C_k}\bigr) \label{eq:color_model}
    \end{align}
    where $\mathbf{L}(x,y) = [L, a, b]^{T}$ is the CIELAB vector. A regularization
    term $\epsilon \mathbf{I}$ ($\epsilon=10^{-3}$) is added to $\boldsymbol{\Sigma}
    _{k}$ for numerical stability.

    \textbf{Mahalanobis distance}: For each pixel in $M_{k}$, compute:
    \begin{equation}
        d_{M}= \sqrt{(\mathbf{L}- \boldsymbol{\mu}_{k})^{T}\boldsymbol{\Sigma}_{k}^{-1}(\mathbf{L}-
        \boldsymbol{\mu}_{k})}\label{eq:mahalanobis}
    \end{equation}
    where we abbreviate $\mathbf{L}= \mathbf{L}(x,y)$ for readability. Pixels
    with $d_{M}> \tau_{M}= 3.5$ are rejected from the mask.

    \subsubsection{Stage~3: Connected Component Cleaning}
    After color validation, small disconnected fragments may remain. We apply
    connected component analysis and remove components smaller than $1\%$ of the
    largest component's area:
    \begin{equation}
        M_{k}^{\text{clean}}= \bigcup_{\{j: |R_j| \geq 0.01 \cdot \max_j |R_j|\}}
        R_{j}\label{eq:cc_clean}
    \end{equation}
    where $R_{j}$ denotes the $j$-th connected region.

    \subsubsection{Stage~4: Cross-Branch Overlap Resolution}
    When multiple masks overlap, each contested pixel is assigned to the branch
    whose Mahalanobis distance is smallest:
    \begin{equation}
        k^{*}(x,y) = \arg\min_{k \in \mathcal{O}(x,y)}d_{M}^{(k)}(x,y) \label{eq:overlap}
    \end{equation}
    where $\mathcal{O}(x,y)$ is the set of branches claiming pixel $(x,y)$.

    \subsection{Version~5: Statistical Depth Optimization}
    \label{sec:v5}

    Once segmentation has been cleaned by Version~4, depth noise within the DEFOM
    disparity map emerges as the dominant remaining artifact. Version~5 combats
    it with a four-phase per-branch depth optimization pipeline.

    \subsubsection{Phase~1: Multi-Round IQR Outlier Removal}
    For each branch $k$, compute the interquartile range of depth values within $M
    _{k}$:
    \begin{align}
        \text{IQR}_{k}          & = Q_{3}- Q_{1}\label{eq:iqr}                                                                            \\
        \text{outlier if }Z_{i} & < Q_{1}- \alpha\!\cdot\!\text{IQR}_{k}\;\text{ or }\; Z_{i}> Q_{3}+ \alpha\!\cdot\!\text{IQR}_{k}\notag
    \end{align}
    with $\alpha = 1.5$. Outliers are replaced with the branch median $\tilde{Z}_{k}$.
    This phase is applied iteratively for up to $T=3$ rounds until no further
    outliers are detected.

    \subsubsection{Phase~2: Z-Score Filtering}
    After IQR clipping, residual outliers are detected via Z-score:
    \begin{equation}
        z_{i}= \frac{|Z_{i}- \bar{Z}_{k}|}{\sigma_{k}}, \quad \text{outlier if }z
        _{i}> 2 \label{eq:zscore}
    \end{equation}
    where $\bar{Z}_{k}$ and $\sigma_{k}$ are the post-IQR branch mean and
    standard deviation.

    \subsubsection{Phase~3: Local Spatial Outlier Detection}
    Using a $7 \times 7$ sliding window, compute the local mean and standard deviation
    via uniform filtering:
    \begin{equation}
        \text{outlier if }|Z_{i}- \tilde{Z}_{i}^{\text{local}}| > \beta \cdot \sigma
        _{i}^{\text{local}}\label{eq:local_outlier}
    \end{equation}
    with $\beta = 2.0$ and $\tilde{Z}_{i}^{\text{local}}$ denoting the local
    median. Detected outliers are replaced with local median values.

    \subsubsection{Phase~4: Spatial Median Filtering}
    A $5 \times 5$ median filter is applied within the mask to smooth residual noise
    while preserving depth edges.

    \textbf{Limitation}: Both IQR and Z-score statistics assume roughly symmetric,
    unimodal distributions---an assumption frequently violated on branch
    surfaces where partial occlusion produces bimodal depth profiles. Moreover,
    the Phase~4 median filter, despite its simplicity, blurs fine-grained depth
    edges. Version~6 redesigns the entire depth pipeline to overcome these shortcomings.

    \subsection{Version~6: Advanced Robust Depth Optimization}
    \label{sec:v6}

    The culminating Version~6 discards the four-phase scheme in favor of a five-stage
    robust optimization, with every stage engineered to address a distinct noise
    characteristic. Table~\ref{tab:depth_stages} provides a concise overview.

    \begin{table}[!t]
        \caption{Five-Stage Depth Optimization Pipeline (Version~6)}
        \label{tab:depth_stages}
        \centering
        \small
        \begin{tabular}{clp{4.2cm}}
            \toprule \textbf{Stage} & \textbf{Method}    & \textbf{Purpose}                  \\
            \midrule 1              & MAD Global         & Robust global outlier detection   \\
            2                       & Spatial Density    & Neighbor consensus voting         \\
            3                       & Local MAD          & Local robust outlier detection    \\
            4                       & Guided Filter      & RGB-guided edge-preserving smooth \\
            5                       & Adaptive Bilateral & Depth + spatial joint weighting   \\
            \bottomrule
        \end{tabular}
    \end{table}

    \subsubsection{Stage~1: MAD Global Robust Outlier Detection}

    The Median Absolute Deviation (MAD)~\cite{leys2013detecting} is a robust
    measure of statistical dispersion with a breakdown point of 50\%, far
    superior to the 25\% breakdown point of IQR. For each branch $k$, compute:
    \begin{equation}
        \text{MAD}_{k}= \text{median}(|Z_{i}- \tilde{Z}_{k}|), \quad i \in M_{k}\label{eq:mad}
    \end{equation}
    The \emph{modified Z-score}~\cite{iglewicz1993detect} for each pixel is:
    \begin{equation}
        m_{i}= \frac{0.6745 \cdot |Z_{i}- \tilde{Z}_{k}|}{\text{MAD}_{k}}\label{eq:modified_z}
    \end{equation}
    where the constant 0.6745 ensures consistency with the standard deviation
    for Gaussian data ($\text{MAD}= 0.6745 \cdot \sigma$ under normality). Pixels
    with $m_{i}> \tau_{\text{MAD}}= 3.5$ are classified as outliers and replaced
    with the branch median $\tilde{Z}_{k}$. This stage is applied iteratively
    for up to $T=3$ rounds.

    \subsubsection{Stage~2: Spatial Density Consensus}

    Global MAD filtering captures extreme outliers but overlooks spatially isolated
    points whose values appear plausible in a branch-wide sense. Stage~2
    introduces neighbor-consensus voting to detect such depth-inconsistent
    pixels:

    Within an $11 \times 11$ window, compute the local median $\tilde{Z}_{i}^{\text{loc}}$
    and local MAD. Convert to a robust local scale estimate:
    \begin{equation}
        \hat{\sigma}_{i}^{\text{loc}}= 1.4826 \cdot \text{MAD}_{i}^{\text{loc}}\label{eq:local_sigma}
    \end{equation}
    where 1.4826 is the consistency constant for a normal distribution. A pixel
    is \emph{consistent} with its neighborhood if:
    \begin{equation}
        |Z_{i}- \tilde{Z}_{i}^{\text{loc}}| \leq \gamma \cdot \hat{\sigma}_{i}^{\text{loc}}
        \label{eq:consistent}
    \end{equation}
    with $\gamma = 2.0$. The \emph{consensus ratio} $\rho_{i}$ is the fraction
    of neighbors satisfying Eq.~\eqref{eq:consistent}. A pixel is classified as an
    isolated outlier if it is itself inconsistent, yet its neighborhood has high
    consensus ($\rho_{i}\geq 0.3$). Such outliers are replaced with $\tilde{Z}_{i}
    ^{\text{loc}}$.

    \subsubsection{Stage~3: Local MAD Outlier Detection}

    Even after global and consensus-based filtering, fine-grained local noise persists.
    Stage~3 re-applies MAD-based detection at the local scale, operating within a
    $7 \times 7$ window:
    \begin{equation}
        |Z_{i}- \tilde{Z}_{i}^{\text{loc}}| > \tau_{L}\cdot 1.4826 \cdot \text{MAD}
        _{i}^{\text{loc}}\label{eq:local_mad}
    \end{equation}
    with $\tau_{L}= 3.0$. Because the local MAD is inherently insensitive to the
    very outliers it seeks to identify, it provides a more reliable scale
    estimate than the standard-deviation approach of Version~5.

    \subsubsection{Stage~4: Guided Filter}

    Edge-preserving smoothing is realized through the guided filter~\cite{he2013guided},
    which leverages the RGB image as a structural reference. Given guidance
    image $I$ and input depth $p$, the output at pixel $i$ is:
    \begin{equation}
        q_{i}= \bar{a}_{k}\cdot I_{i}+ \bar{b}_{k}\label{eq:guided_filter}
    \end{equation}
    where $\bar{a}_{k}$ and $\bar{b}_{k}$ are averaged linear coefficients
    computed as:
    \begin{align}
        a_{k} & = \frac{\frac{1}{|\omega_k|}\sum_{i \in \omega_k}I_{i}p_{i}- \bar{I}_{k}\bar{p}_{k}}{\sigma_{k}^{2}+ \epsilon}\label{eq:guided_coeff} \\
        b_{k} & = \bar{p}_{k}- a_{k}\bar{I}_{k}\notag
    \end{align}
    with window $\omega_{k}$ of radius $r=4$ and regularization $\epsilon = 0.01$.

    We construct a \emph{multi-channel guidance} signal by combining grayscale
    intensity, and CIELAB L, a, b channels with weights $[0.4, 0.3, 0.15, 0.15]$,
    ensuring that both luminance and chrominance edges are respected during
    depth smoothing.

    \subsubsection{Stage~5: Adaptive Bilateral Filtering}

    A final bilateral filtering pass combines spatial proximity and depth similarity
    into a joint weighting scheme with \emph{data-adaptive} parameters. The filtered
    depth at pixel $i$ is:
    \begin{equation}
        q_{i}= \frac{\sum_{j \in \mathcal{N}_i}w_{s}(i,j) \cdot w_{d}(i,j) \cdot
        Z_{j}}{\sum_{j \in \mathcal{N}_i}w_{s}(i,j) \cdot w_{d}(i,j)}\label{eq:bilateral}
    \end{equation}
    where $w_{s}(i,j) = \exp(-\|p_{i}- p_{j}\|^{2}/ 2\sigma_{s}^{2})$ is the
    spatial weight and $w_{d}(i,j) = \exp(-|Z_{i}- Z_{j}|^{2}/ 2\sigma_{d}^{2})$
    is the depth-range weight.

    Critically, the depth-range bandwidth $\sigma_{d}$ is automatically tuned per
    branch using the branch's own MAD:
    \begin{equation}
        \sigma_{d}= \alpha_{d}\cdot 1.4826 \cdot \text{MAD}_{k}\label{eq:adaptive_sigma}
    \end{equation}
    with $\alpha_{d}= 1.5$ and spatial bandwidth $\sigma_{s}= 7$~px. This adaptation
    ensures that branches with larger depth variation (e.g., trunks) receive broader
    smoothing, while thin branches with tight depth distributions receive more
    conservative filtering.

    \begin{figure*}[!t]
        \centering
        \includegraphics[width=\textwidth]{
            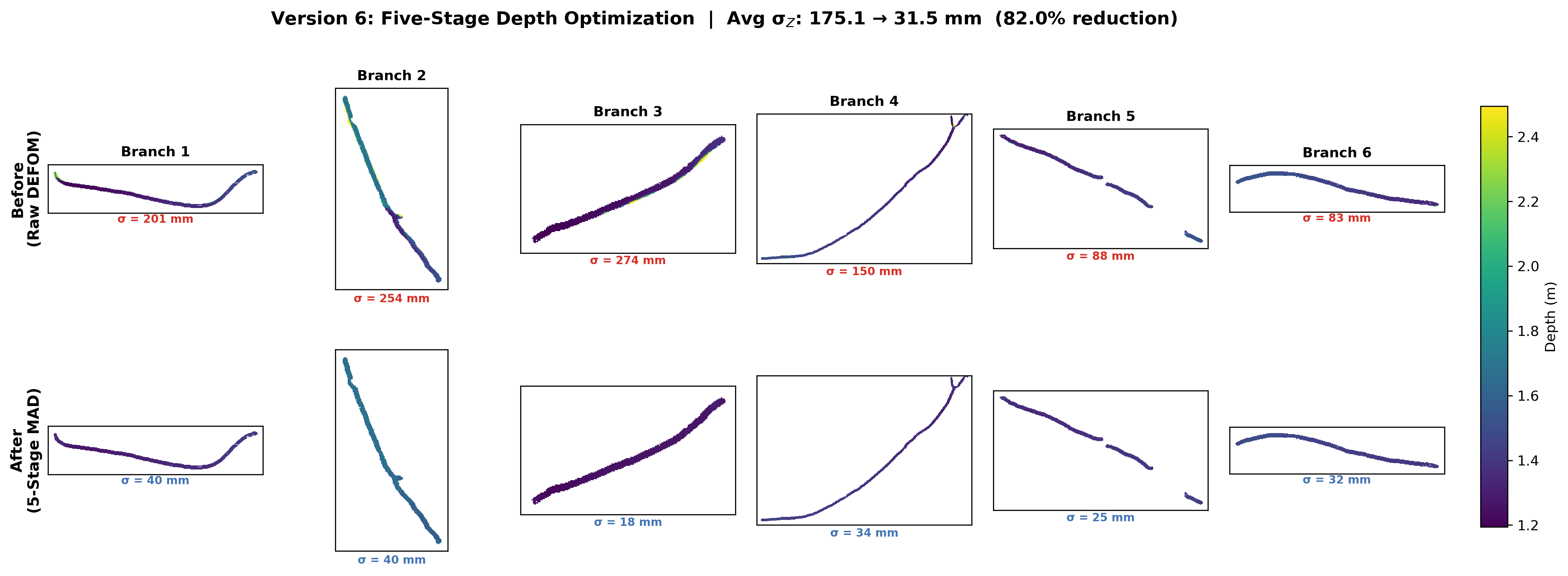
        }
        \caption{Version~6 five-stage depth optimization results. Top row:
        before optimization (V4 masks with raw DEFOM depth). Bottom row: after
        the five-stage MAD-based pipeline. Average $\sigma_{Z}$ reduced from 174.6~mm
        to 31.5~mm (82.0\% reduction).}
        \label{fig:v6_stages}
    \end{figure*}

    \section{Experimental Results}

    \subsection{Experimental Setup}

    \textbf{Hardware}: All experiments run on Google Colab with NVIDIA A100/V100
    GPU. DEFOM-Stereo uses a ViT-L DINOv2 backbone with 32 valid iterations.

    \textbf{Data}: We process stereo pairs of Radiata pine branches at
    $1920 \times 1080$ resolution captured with a ZED~Mini camera (focal length $f
    _{x}\approx 1120$~px, baseline $B = 63$~mm).

    \textbf{Evaluation metrics}: We evaluate per-branch depth quality using:
    \begin{itemize}
        \item \textbf{Depth standard deviation} ($\sigma_{Z}$): lower values indicate
            less noise within a single branch.

        \item \textbf{Depth range} (max$-$min): smaller ranges indicate fewer extreme
            outliers and tighter per-branch distributions.

        \item \textbf{Total mask pixels}: retained mask area; a drop indicates
            loss of thin structures.

        \item \textbf{3D point cloud coherence}: qualitatively assessed via per-branch
            colored point clouds and depth histograms.
    \end{itemize}

    \subsection{Version-by-Version Comparison}

\begin{figure}[!t]
        \centering
        \includegraphics[width=\columnwidth]{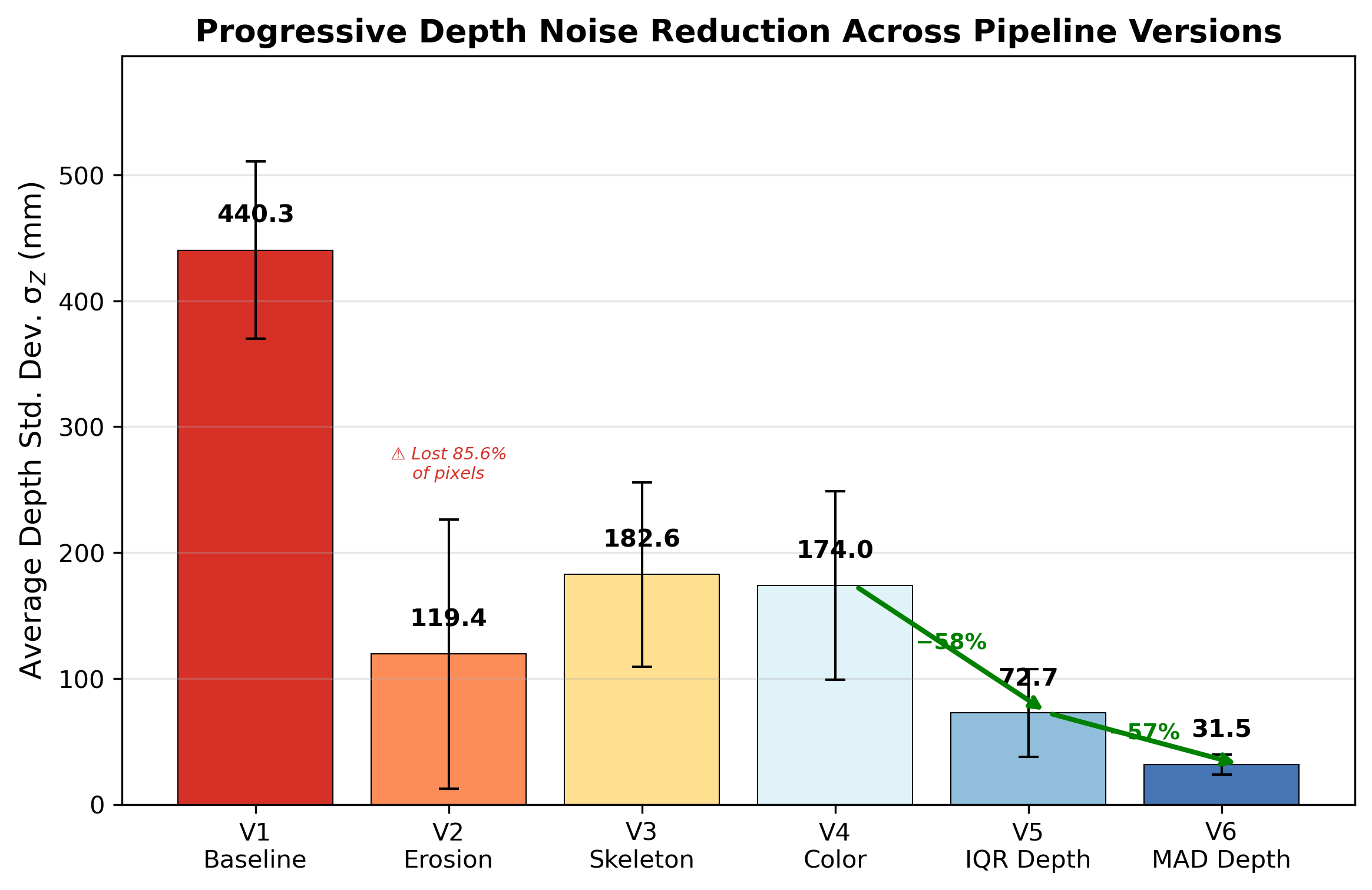}
        \caption{Average per-branch depth standard deviation across pipeline
        versions. V2 achieves low $\sigma_{Z}$ but loses 85.6\% of mask pixels. V6
        achieves the lowest $\sigma_{Z}$ (31.5~mm) while preserving all branches.}
        \label{fig:std_barchart}
    \end{figure}

    Table~\ref{tab:version_comparison} summarizes the progressive improvement
    across all six versions. Fig.~\ref{fig:std_barchart} visualizes the average
    depth standard deviation reduction.

    \begin{table}[!t]
        \caption{Progressive Improvement Across Pipeline Versions}
        \label{tab:version_comparison}
        \centering
        \resizebox{\columnwidth}{!}{
        \begin{tabular}{lccccc}
            \toprule \textbf{Ver.} & \textbf{Avg.~$\sigma_{Z}$ (mm)} & \textbf{Range (mm)} & \textbf{Pixels} & \textbf{Mask Qual.} & \textbf{Thin Br.} \\
            \midrule V1 (Baseline) & 440.3                           & 1545                & 27\,318         & Low                 & N/A               \\
            V2 (Erosion)           & 119.4                           & 461                 & 3\,935          & Medium              & No                \\
            V3 (Skeleton)          & 182.6                           & 1431                & 14\,960         & Medium              & Yes               \\
            V4 (Color)             & 174.0                           & 1431                & 14\,653         & High                & Yes               \\
            V5 (IQR)               & 72.7                            & 281                 & 14\,653         & High                & Yes               \\
            V6 (MAD)               & \textbf{31.5}                   & \textbf{128}        & 14\,665         & High                & Yes               \\
            \bottomrule
        \end{tabular}}
        \vspace{0.5em}
        \begin{flushleft}
            \footnotesize{Avg.~$\sigma_{Z}$ and Avg.~Range are averaged across all six branches. Total Pixels is the sum of valid mask pixels across branches. V2 loses 85.6\% of pixels due to na\"{i}ve erosion. V5 and V6 share V4's masks but apply depth optimization.}
        \end{flushleft}
    \end{table}

    \subsection{Mask Refinement Analysis (V1--V4)}

    \begin{figure}[!t]
        \centering
        \includegraphics[width=\columnwidth]{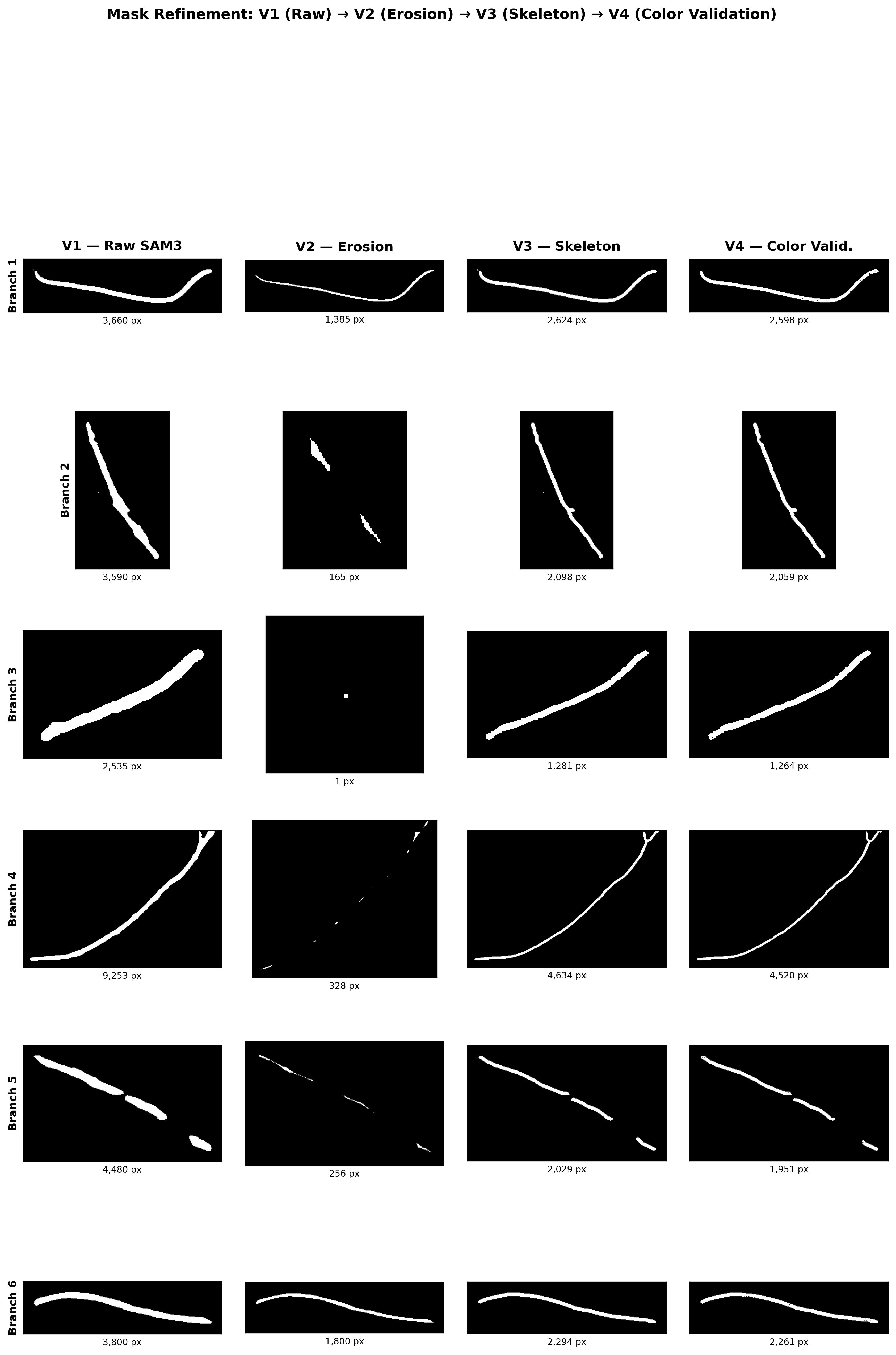}
        \caption{Mask refinement progression from V1 to V4. V2 erosion disconnects
        thin branches (Branch~3: 2\,535$\to$1~px). V3 skeleton preservation recovers
        connectivity (1\,281~px). V4 color validation produces clean masks.}
        \label{fig:mask_comparison}
    \end{figure}

    \textbf{V1 $\rightarrow$ V2}: The mask refinement progression from V1 to V4
    is laid out in Fig.~\ref{fig:mask_comparison}. Shrinking each mask inward by
    15~px strips away the majority of sky-pixel outliers from depth histograms. The
    cost, however, is severe: thin branches whose diameter falls below 30~px are
    disconnected or lost entirely.

    \textbf{V2 $\rightarrow$ V3}: Skeleton-preserving erosion recovers this lost
    connectivity. By extracting the topological centerline and dilating it to $r_{s}
    = \max(3, r_{e}/4)$~pixels before merging with the distance-eroded mask, the
    algorithm eliminates topology loss while retaining boundary erosion for
    larger structures.

    \textbf{V3 $\rightarrow$ V4}: Geometric refinement alone cannot catch pixels
    that lie within the mask boundary yet differ in color from the branch. The LAB
    Mahalanobis color model (Eq.~\ref{eq:color_model}) built from the deeply-eroded
    core provides a robust reference, and the $\tau_{M}= 3.5$ threshold strikes a
    balance between tolerating natural color variability and excising contaminants.
    Subsequent cross-branch overlap resolution (Eq.~\ref{eq:overlap}) produces the
    non-overlapping masks required for unbiased per-branch depth statistics.

    \subsection{Depth Optimization Analysis (V5--V6)}

    \begin{figure}[!t]
        \centering
        \includegraphics[width=\columnwidth]{
            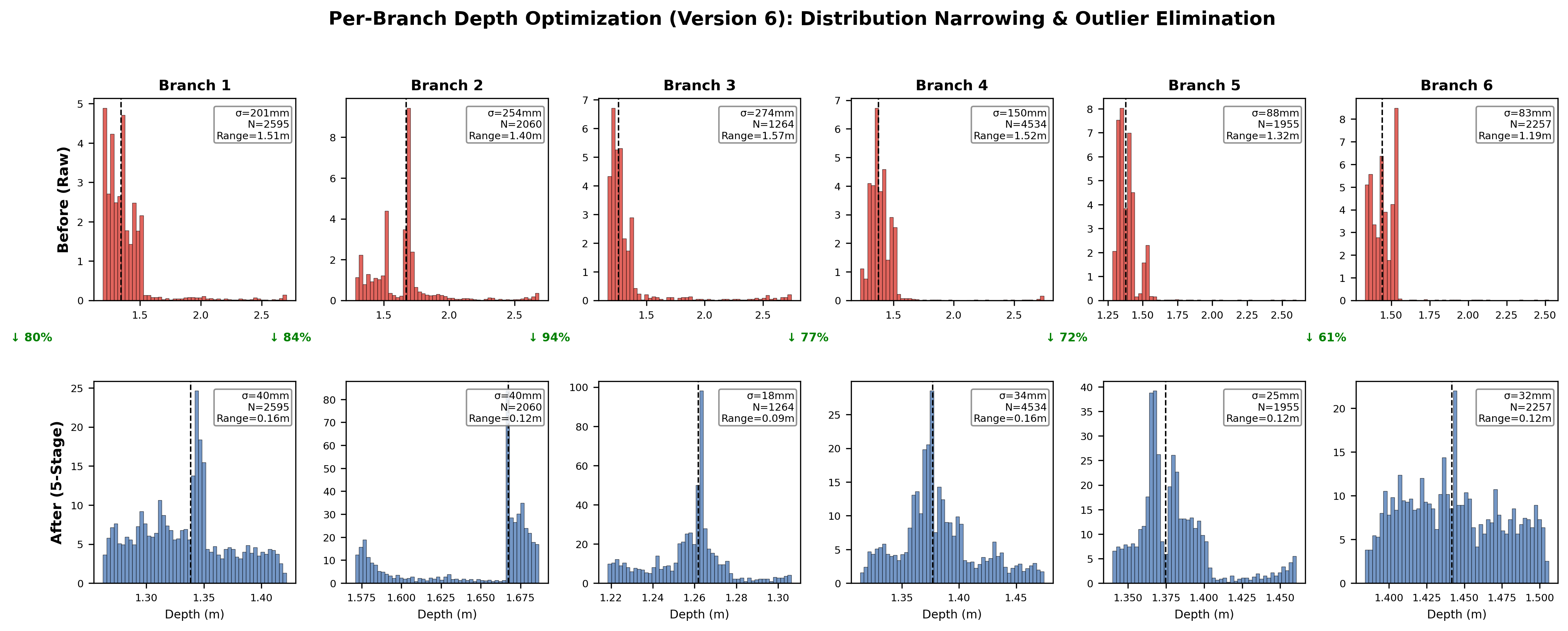
        }
        \caption{Per-branch depth optimization comparison (Version~6). Before (top)
        vs.\ after (bottom) the five-stage pipeline, showing distribution narrowing
        and outlier elimination for each branch.}
        \label{fig:depth_histograms}
    \end{figure}

    Per-branch depth distributions before and after the Version~6 optimization are
    contrasted in Fig.~\ref{fig:depth_histograms}.

    \textbf{V4 $\rightarrow$ V5}: Applying the four-phase IQR-based pipeline
    yields a marked reduction in depth noise. Gross outliers are clipped by multi-round
    IQR; residual large deviations fall to Z-score filtering; isolated local noise
    is caught by spatial detection; and a final median pass smooths what remains.
    A notable drawback is that the $5 \times 5$ median kernel blurs depth edges,
    and the 25\% breakdown point of IQR means that extreme outliers can still inflate
    $Q_{1}/Q_{3}$ enough to escape detection.

    \textbf{V5 $\rightarrow$ V6}: The five-stage MAD-based pipeline (Fig.~\ref{fig:v6_stages})
    overcomes each of these limitations:

    \begin{enumerate}
        \item \textbf{MAD vs.\ IQR}: With a 50\% breakdown point, MAD can tolerate
            outliers constituting up to half the data, far exceeding IQR's 25\% ceiling.

        \item \textbf{Consensus voting vs.\ Z-score}: Neighborhood agreement captures
            spatially isolated anomalies that global statistics overlook.

        \item \textbf{Guided filter vs.\ median filter}: By respecting RGB-aligned
            depth discontinuities, the guided filter retains sharper branch
            boundaries in 3D.

        \item \textbf{Adaptive bilateral smoothing}: Branch-specific $\sigma_{d}$
            tuning (Eq.~\ref{eq:adaptive_sigma}) scales the smoothing bandwidth
            to each branch's depth variability, preventing over-smoothing of narrow
            branches and under-smoothing of wide trunks.
    \end{enumerate}

    \subsection{3D Point Cloud Quality}

    \begin{figure*}[!t]
        \centering
        \includegraphics[width=\textwidth]{
            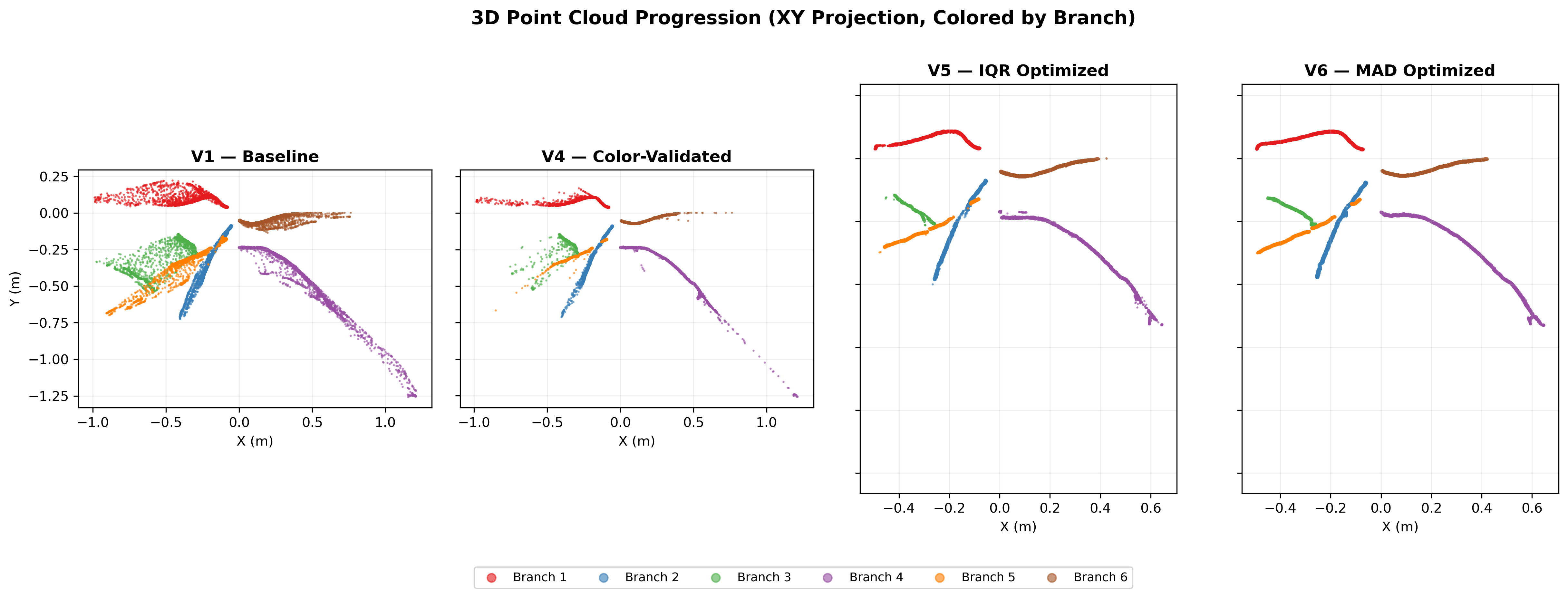
        }
        \caption{Per-branch 3D point cloud progression from V1 to V6. Version~1 exhibits
        severe scatter from mask contamination and depth noise. Version~4
        removes mask-related scatter. Version~6 achieves tight depth clustering with
        minimal outliers.}
        \label{fig:3d_comparison}
    \end{figure*}

\begin{figure}[!t]
        \centering
        \includegraphics[width=\columnwidth]{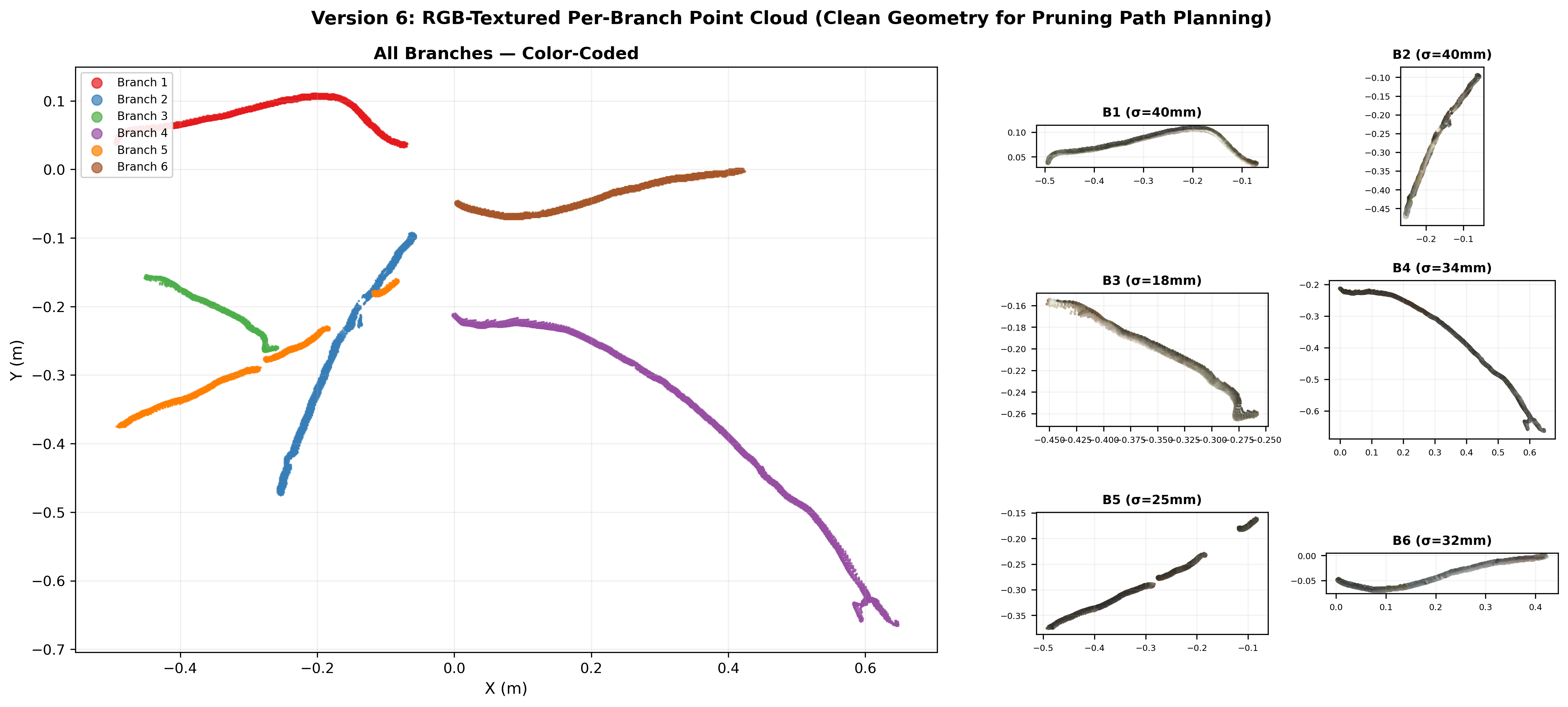}
        \caption{Final Version~6 RGB-textured per-branch point cloud. Each branch
        is color-coded; clean geometry is suitable for autonomous pruning path
        planning.}
        \label{fig:interactive_3d}
    \end{figure}

    Across pipeline versions, 3D point cloud fidelity improves substantially (Fig.~\ref{fig:3d_comparison}).
    Version~1 clouds are dominated by scatter originating from both mask
    contamination and depth noise. Mask-related scatter disappears by Version~4,
    though depth-noise scatter persists. Version~5 curbs depth noise at the expense
    of edge blur. Version~6 strikes the most favorable balance: depth values cluster
    tightly within each branch, boundaries remain sharp where they coincide with
    RGB edges, and isolated outlier points are nearly absent. The final RGB-textured
    per-branch point cloud is shown in Fig.~\ref{fig:interactive_3d}.

    \subsection{Per-Branch Depth Distribution Analysis}

    \begin{figure}[!t]
        \centering
        \includegraphics[width=\columnwidth]{
            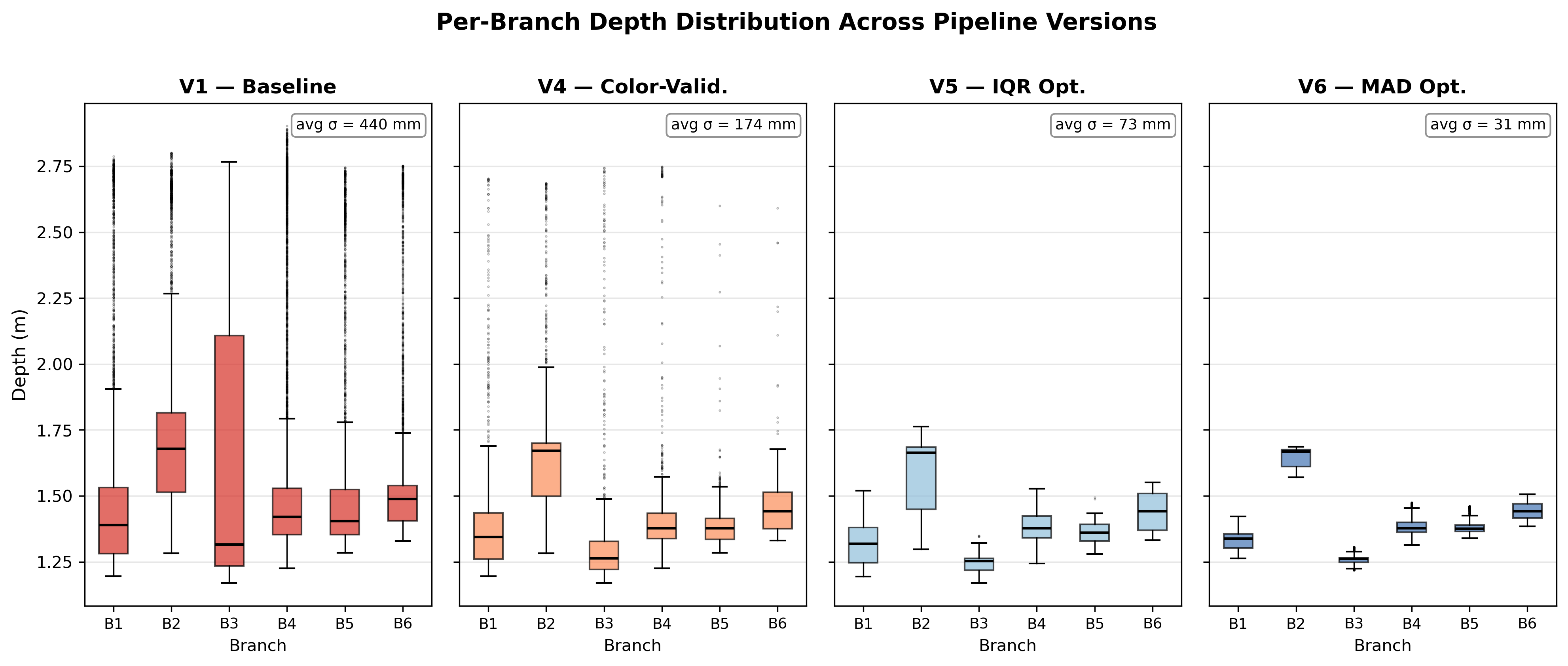
        }
        \caption{Box plots comparing per-branch depth distributions across
        pipeline versions. Version~6 shows the tightest distributions with the fewest
        outliers.}
        \label{fig:boxplots}
    \end{figure}

    Box plots (Fig.~\ref{fig:boxplots}) expose the cumulative impact of each refinement
    on per-branch depth distributions. Version~1 exhibits extreme outliers---driven
    by sky-depth contamination---and wide IQR ranges. Successive mask refinements
    (V2--V4) progressively purge distribution contamination, while the depth optimization
    stages (V5--V6) further compress the IQR and eliminate residual whiskers, ultimately
    yielding per-branch depth estimates precise enough for centimeter-level tool
    positioning.

    \subsection{Ablation Study: Depth Optimization Stages}

    To isolate the contribution of depth optimization from mask refinement, Table~\ref{tab:ablation_stages}
    compares Version~5 (IQR-based) and Version~6 (MAD-based) starting from identical
    V4-refined masks.

    \begin{table}[!t]
        \caption{Per-Branch Depth Optimization: V5 vs.\ V6}
        \label{tab:ablation_stages}
        \centering
        \resizebox{\columnwidth}{!}{
        \begin{tabular}{lrrrr}
            \toprule \textbf{Branch}  & \textbf{Before $\sigma_{Z}$} & \textbf{V5 $\sigma_{Z}$} & \textbf{V6 $\sigma_{Z}$} & \textbf{V6 Red.} \\
                                      & \textbf{(mm)}                & \textbf{(mm)}            & \textbf{(mm)}            & \textbf{(\%)}    \\
            \midrule B1 ($N$=2\,598)  & 201.7                        & 91.7                     & \textbf{39.7}            & 80.3             \\
            B2 ($N$=2\,059)           & 252.7                        & 137.9                    & \textbf{39.9}            & 84.2             \\
            B3 ($N$=1\,264)           & 274.9                        & 35.4                     & \textbf{17.6}            & 93.6             \\
            B4 ($N$=4\,520)           & 146.9                        & 65.3                     & \textbf{34.4}            & 76.6             \\
            B5 ($N$=1\,951)           & 84.5                         & 37.6                     & \textbf{24.9}            & 70.5             \\
            B6 ($N$=2\,261)           & 86.9                         & 68.3                     & \textbf{32.4}            & 62.7             \\
            \midrule \textbf{Average} & 174.6                        & 72.7                     & \textbf{31.5}            & \textbf{82.0}    \\
            \bottomrule
        \end{tabular}}
        \vspace{0.5em}
        \begin{flushleft}
            \footnotesize{``Before'' is the V4-refined mask without depth optimization. V5 uses IQR+Z-score+median filter (1\,738 pixels modified). V6 uses MAD+spatial consensus+guided+bilateral filter (4\,623 pixels modified). Reduction computed relative to Before.}
        \end{flushleft}
    \end{table}

    \section{Discussion}

    \subsection{Rationale for Progressive Refinement}

    The six-version arc is not mere incremental tinkering; it is a diagnostic sequence
    in which every version exposes a distinct failure mode that shapes the subsequent
    design decision:

    \begin{itemize}
        \item Version~1 uncovered the inadequacy of SAM3's boundary precision for
            depth-sensitive tasks, pointing to the need for mask erosion.

        \item The thin-branch topology destruction caused by na\"ive erosion in Version~2
            prompted the skeleton-preserving strategy of Version~3.

        \item Version~3 made clear that geometry-only mask refinement leaves color-inconsistent
            pixels intact, necessitating Mahalanobis color validation.

        \item With mask quality secured, Version~4 exposed depth noise as the
            single largest obstacle to accurate per-branch reconstruction.

        \item Version~5's reliance on IQR and Z-score proved fragile under the heavy-tailed
            noise distributions typical of vegetation stereo, driving the
            redesign around MAD-based statistics.
    \end{itemize}

    This chain of diagnoses guarantees that every component is both \emph{necessary}---removing
    it measurably degrades output quality---and \emph{targeted at a genuinely
    different failure mode}.

    \subsection{Robustness of MAD vs.\ IQR}

    Switching from IQR-based (Version~5) to MAD-based (Version~6 Stage~1) outlier
    detection rests on a clear theoretical foundation: the \emph{breakdown point}---the
    largest fraction of outliers a statistic can tolerate before its estimate becomes
    arbitrarily unreliable. IQR breaks down at 25\%: once more than a quarter of
    depth values are corrupted, $Q_{1}$ and $Q_{3}$ themselves become distorted.
    MAD, by contrast, withstands contamination of up to 50\% of the data. In
    forestry imagery, branches frequently occupy only a minor share of the mask area---particularly
    after partial occlusion---making the doubled resilience of MAD a decisive practical
    advantage.

    \subsection{Edge Preservation: Guided Filter vs.\ Median Filter}

    Replacing the median filter (Version~5) with an RGB-guided filter (Version~6
    Stage~4) tackles a core deficiency: median filtering is spatially isotropic and
    blind to image structure. The guided filter, in contrast, borrows edge
    information from the color image, preserving depth discontinuities wherever they
    align with color transitions. For tree branches---where the depth boundary
    and the branch-background color boundary are essentially co-located---this
    property is especially valuable.

    \subsection{Practical Implications}

    Beyond noise reduction, the Version~6 pipeline unlocks several downstream
    capabilities critical for autonomous pruning:

    \begin{itemize}
        \item \textbf{Branch diameter estimation}: Tighter depth distributions support
            reliable depth-profile analysis for measuring branch cross-sections.

        \item \textbf{Pruning path planning}: Geometrically clean per-branch point
            clouds furnish the spatial input needed by cutting-tool trajectory optimizers.

        \item \textbf{Distance estimation}: Narrow per-branch depth distributions
            translate directly into precise UAV-to-branch distance readings during
            approach maneuvers.
    \end{itemize}

    \subsection{Limitations}

    \textbf{Pseudo-Ground-Truth Ceiling.} Depth accuracy is fundamentally capped
    by DEFOM-Stereo's disparity estimation. Prediction errors in textureless sky
    regions or heavily occluded areas propagate through every subsequent
    optimization stage.

    \textbf{Parameter Sensitivity.} The pipeline relies on several hand-tuned parameters
    ($\tau_{M}$, $\tau_{\text{MAD}}$, $r_{e}$, $r_{c}$, $\epsilon$, $\alpha_{d}$).
    Although the chosen values generalize well within our test scenes, different
    tree species or camera configurations may demand re-tuning.

    \textbf{Computational Cost.} DEFOM-Stereo inference dominates the overall runtime;
    the added mask refinement and depth optimization stages contribute
    negligible overhead.

    \textbf{Limited Scene Diversity.} All experiments involve Radiata pine in Canterbury,
    New Zealand. Broader validation across species, seasons, and lighting conditions
    remains necessary.

    \section{Conclusion}

    This paper introduced a progressive pipeline that unites DEFOM-Stereo
    disparity estimation, SAM3 instance segmentation, and multi-stage depth
    optimization for per-branch 3D reconstruction in UAV forestry. Over six successive
    versions, three distinct error families were identified and resolved: mask boundary
    contamination through skeleton-preserving erosion, segmentation inaccuracy
    through LAB Mahalanobis color validation, and depth noise through a five-stage
    MAD-based robust optimization augmented by guided and adaptive bilateral filtering.

    The following findings stand out:

    \begin{itemize}
        \item \textbf{Skeleton-preserving erosion} suppresses boundary
            contamination without sacrificing thin-branch connectivity, surpassing
            conventional morphological erosion.

        \item \textbf{LAB Mahalanobis color validation} effectively purges color-inconsistent
            pixels and arbitrates cross-branch overlaps, yielding clean, non-overlapping
            instance masks.

        \item \textbf{MAD-based outlier detection}, with its 50\% breakdown
            point, proves substantially more reliable than IQR-based methods (25\%
            breakdown) under the heavy-tailed noise typical of vegetation depth
            data.

        \item \textbf{Multi-channel guided filtering} respects depth
            discontinuities aligned with RGB edges, delivering sharper branch reconstructions
            than isotropic median filtering.

        \item \textbf{Adaptive bilateral filtering} tunes $\sigma_{d}$ per
            branch, matching smoothing intensity to each branch's intrinsic depth
            variability.
    \end{itemize}

    Taken together, these refinements produce geometrically coherent per-branch 3D
    point clouds suitable for pruning path planning, branch diameter measurement,
    and real-time distance computation. Code, processed outputs, and interactive
    3D visualizations are publicly released. Looking ahead, we plan to couple
    this pipeline with branch detection~\cite{lin2024branch} and segmentation~\cite{lin2025segmentation}
    systems toward fully autonomous pruning, and to explore temporal consistency
    constraints enabling video-rate per-branch tracking.

    \section*{Acknowledgments}
    This research was supported by the Royal Society of New Zealand Marsden Fund
    and the Ministry of Business, Innovation and Employment. We thank the forestry
    research stations for data collection access.

    \bibliographystyle{IEEEtran}

\begin{thebibliography}{99}
        \bibitem{lin2024branch} Y. Lin, B. Xue, M. Zhang, S. Schofield, and R.
            Green, ``Deep learning-based depth map generation and YOLO-integrated
            distance estimation for radiata pine branch detection using drone stereo
            vision,'' in \textit{Proc. Int. Conf. Image Vis. Comput. New Zealand
            (IVCNZ)}, Christchurch, New Zealand, 2024, pp. 1--6.

        \bibitem{steininger2025timbervision} D. Steininger, J. Simon, A. Trondl,
            and M. Murschitz, ``TimberVision: A multi-task dataset and framework
            for log-component segmentation and tracking in autonomous forestry
            operations,'' in \textit{Proc. IEEE/CVF Winter Conf. Appl. Comput.
            Vis. (WACV)}, 2025.

        \bibitem{lin2024benchmark} Y. Lin, B. Xue, M. Zhang, S. Schofield, and R.
            Green, ``Towards gold-standard depth estimation for tree branches in
            UAV forestry: Benchmarking deep stereo matching methods,'' \textit{arXiv
            preprint arXiv:2601.19461}, 2026.

        \bibitem{jiang2025defom} H. Jiang, Z. Lou, L. Ding, R. Xu, M. Tan, W.
            Jiang, and R. Huang, ``DEFOM-Stereo: Depth foundation model based
            stereo matching,'' \textit{arXiv preprint arXiv:2501.09466}, 2025.

        \bibitem{ravi2024sam2} N. Ravi, V. Gabeur, Y.-T. Hu, R. Hu, C. Ryali, T.
            Ma, H. Khedr, R. R\"{a}dle, C. Rolland, L. Gustafson, E. Mintun, J.
            Pan, K. V. Alwala, N. Carion, C.-Y. Wu, R. Girshick, P. Doll\'{a}r,
            and C. Feichtenhofer, ``SAM 2: Segment anything in images and videos,''
            \textit{arXiv preprint arXiv:2408.00714}, 2024.

        \bibitem{kirillov2023sam} A. Kirillov, E. Mintun, N. Ravi, H. Mao, C.
            Rolland, L. Gustafson, T. Xiao, S. Whitehead, A. C. Berg, W.-Y. Lo, P.
            Doll\'{a}r, and R. Girshick, ``Segment anything,'' in \textit{Proc. IEEE
            Int. Conf. Comput. Vis. (ICCV)}, 2023, pp. 4015--4026.

        \bibitem{chang2018psmnet} J.-R. Chang and Y.-S. Chen, ``Pyramid stereo
            matching network,'' in \textit{Proc. IEEE Conf. Comput. Vis. Pattern
            Recognit. (CVPR)}, 2018, pp. 5410--5418.

        \bibitem{guo2019gwcnet} X. Guo, K. Yang, W. Yang, X. Wang, and H. Li, ``Group-wise
            correlation stereo network,'' in \textit{Proc. IEEE Conf. Comput. Vis.
            Pattern Recognit. (CVPR)}, 2019, pp. 3273--3282.

        \bibitem{lipson2021raft} L. Lipson, Z. Teed, and J. Deng, ``RAFT-Stereo:
            Multilevel recurrent field transforms for stereo matching,'' in \textit{Proc.
            Int. Conf. 3D Vis. (3DV)}, 2021, pp. 218--227.

        \bibitem{oquab2024dinov2} M. Oquab, T. Darcet, T. Moutakanni, H. Vo, M.
            Szafraniec, V. Khalidov, P. Fernandez, D. Haziza, F. Massa, A. El-Nouby,
            M. Assran, N. Ballas, W. Galuba, R. Howes, P.-Y. Huang, S.-W. Li, I.
            Misra, M. Rabbat, V. Sharma, G. Synnaeve, H. Xu, H. Jegou, J. Mairal,
            P. Labatut, A. Joulin, and P. Bojanowski, ``DINOv2: Learning robust
            visual features without supervision,'' \textit{Trans. Mach. Learn. Res.},
            2024.

        \bibitem{schops2017eth3d} T. Sch{\"o}ps, J. L. Sch{\"o}nberger, S. Galliani,
            T. Sattler, K. Schindler, M. Pollefeys, and A. Geiger, ``A multi-view
            stereo benchmark with high-resolution images and multi-camera videos,''
            in \textit{Proc. IEEE Conf. Comput. Vis. Pattern Recognit. (CVPR)}, 2017,
            pp. 2538--2547.

        \bibitem{geiger2012kitti} A. Geiger, P. Lenz, and R. Urtasun, ``Are we ready
            for autonomous driving? The KITTI vision benchmark suite,'' in \textit{Proc.
            IEEE Conf. Comput. Vis. Pattern Recognit. (CVPR)}, 2012, pp. 3354--3361.

        \bibitem{scharstein2014middlebury} D. Scharstein, H. Hirschm{\"u}ller, Y.
            Kitajima, G. Krathwohl, N. Ne{\v{s}}i{\'c}, X. Wang, and P. Westling,
            ``High-resolution stereo datasets with subpixel-accurate ground
            truth,'' in \textit{Proc. German Conf. Pattern Recognit. (GCPR)}, 2014,
            pp. 31--42.

        \bibitem{mayer2016large} N. Mayer, E. Ilg, P. Hausser, P. Fischer, D.
            Cremers, A. Dosovitskiy, and T. Brox, ``A large dataset to train convolutional
            networks for disparity, optical flow, and scene flow estimation,''
            in \textit{Proc. IEEE Conf. Comput. Vis. Pattern Recognit. (CVPR)},
            2016, pp. 4040--4048.

        \bibitem{he2013guided} K. He, J. Sun, and X. Tang, ``Guided image filtering,''
            \textit{IEEE Trans. Pattern Anal. Mach. Intell.}, vol.~35, no.~6, pp.
            1397--1409, Jun. 2013.

        \bibitem{tomasi1998bilateral} C. Tomasi and R. Manduchi, ``Bilateral
            filtering for gray and color images,'' in \textit{Proc. IEEE Int.
            Conf. Comput. Vis. (ICCV)}, 1998, pp. 839--846.

        \bibitem{ma2013constant} Z. Ma, K. He, Y. Wei, J. Sun, and E. Wu, ``Constant
            time weighted median filtering for stereo matching and beyond,'' in
            \textit{Proc. IEEE Int. Conf. Comput. Vis. (ICCV)}, 2013, pp. 49--56.

        \bibitem{leys2013detecting} C. Leys, C. Ley, O. Klein, P. Bernard, and L.
            Licata, ``Detecting outliers: Do not use standard deviation around
            the mean, use absolute deviation around the median,'' \textit{J. Exp.
            Social Psych.}, vol.~49, no.~4, pp. 764--766, 2013.

        \bibitem{iglewicz1993detect} B. Iglewicz and D. C. Hoaglin, \textit{Volume
            16: How to Detect and Handle Outliers}.\hskip 1em plus 0.5em minus 0.4em\relax
            Milwaukee, WI: ASQC Quality Press, 1993.

        \bibitem{lee1994skeleton} T.-C. Lee, R. L. Kashyap, and C.-N. Chu, ``Building
            skeleton models via 3-D medial surface/axis thinning algorithms,'' \textit{CVGIP:
            Graph. Models Image Process.}, vol.~56, no.~6, pp. 462--478, 1994.

        \bibitem{fraundorfer2012vision} F. Fraundorfer, L. Heng, D. Honegger, G.
            H. Lee, L. Meier, P. Tanskanen, and M. Pollefeys, ``Vision-based
            autonomous mapping and exploration using a quadrotor MAV,'' in \textit{Proc.
            IEEE/RSJ Int. Conf. Intell. Robots Syst. (IROS)}, 2012, pp. 4557--4564.

        \bibitem{nex2014uav} F. Nex and F. Remondino, ``UAV for 3D mapping
            applications: A review,'' \textit{Appl. Geomat.}, vol.~6, no.~1, pp.
            1--15, 2014.

        \bibitem{barry2015pushbroom} A. J. Barry and R. Tedrake, ``Pushbroom stereo
            for high-speed navigation in cluttered environments,'' in \textit{Proc.
            IEEE Int. Conf. Robot. Autom. (ICRA)}, 2015, pp. 3046--3052.

        \bibitem{lin2025yolosgbm} Y. Lin, B. Xue, M. Zhang, S. Schofield, and R.
            Green, ``YOLO and SGBM integration for autonomous tree branch detection
            and depth estimation in radiata pine pruning applications,'' in
            \textit{Proc. Int. Conf. Image Vis. Comput. New Zealand (IVCNZ)},
            Wellington, New Zealand, 2025, pp. 1--6.

        \bibitem{lin2025segmentation} Y. Lin, B. Xue, M. Zhang, S. Schofield, and
            R. Green, ``Performance evaluation of deep learning for tree branch segmentation
            in autonomous forestry systems,'' in \textit{Proc. Int. Conf. Image Vis.
            Comput. New Zealand (IVCNZ)}, Wellington, New Zealand, 2025, pp. 1--6.
    \end{thebibliography}
    
\end{document}